\documentclass[12pt]{article}

\usepackage{graphicx}
\usepackage{siunitx}
\usepackage{xcolor}
\usepackage[export]{adjustbox}
\usepackage{times}
\usepackage{dblfloatfix}

\newenvironment{sciabstract}{%
\begin{quote} \bf Machine learning software applications are nowadays ubiquitous in many fields of science and society for their outstanding capability of solving computationally vast problems like the recognition of patterns and regularities in big datasets. One of the main goals of research is the realization of a physical neural network able to perform data processing in a much faster and energy-efficient way than the state-of-the-art technology. Here we show that lattices of exciton-polariton condensates accomplish neuromorphic computing using fast optical nonlinearities and with lower error rate than any previous hardware implementation. We demonstrate that our neural network significantly increases the recognition efficiency compared to the linear classification algorithms on one of the most widely used benchmarks, the MNIST problem, showing a concrete advantage from the integration of optical systems in reservoir computing architectures.}
{\end{quote}}

\title{Polaritonic neuromorphic computing outperforms linear classifiers}

\author
{D. Ballarini,$^{1}$ A. Gianfrate,$^{1}$ R. Panico,$^{1}$ A. Opala,$^{2}$\\ S. Ghosh,$^{3}$ L. Dominici,$^{1}$ V. Ardizzone,$^{1}$ M. De Giorgi,$^{1}$ G. Lerario,$^{1}$ G. Gigli,$^{1}$\\ T.C.H. Liew,$^{3}$  M. Matuszewski,$^{2}$ D. Sanvitto,$^{1}$\\
\normalsize{$^{1}$CNR NANOTEC--Institute of Nanotechnology, Via Monteroni, 73100 Lecce, Italy}\\
\normalsize{$^{2}$Institute of Physics, Polish Academy of Sciences,}\\ \normalsize{Al. Lotników 32/46, PL-02-668 Warsaw, Poland}\\
\normalsize{$^{3}$School of Physical and Mathematical Science,} \\ \normalsize{Nanyang Technological University 637371, Singapore}}

\date{}

\begin{document}

\baselineskip24pt

\maketitle

\begin{sciabstract}
 
\end{sciabstract}

\paragraph*{Introduction}

Artificial neural networks emerged as powerful information processing systems for the recognition and classification of patterns even with imperfect, noisy, or incomplete information. 
However, software implementations suffer from the so-called von Neumann bottleneck due to the physical separation of memory and processing units. On the other hand, neuromorphic designs attempt to adjust the structure of a physical system to the architecture of the neural network. This allows a native parallelism of tasks, increasing the efficiency of the computational process and avoiding the bottleneck of the von Neumann architecture.

Aside implementations using conventional binary logic to simulate neural networks, hardware implementations~\cite{Misra2010} have been based on memristors~\cite{Serb2016, Hu2015, Prezioso2015}, spintronic~\cite{Sengputa2016}, and optical systems~\cite{Yichen2017}. 
Yet, the individual control of the weights of connections between nodes places a challenging requirement on hardware implementations of neural networks. Moreover, even being able to independently control many weight connections, one still has the task of choosing them. 
A promising solution appeared in the form of reservoir computing (RC)~\cite{Maas2002, Herbert2004}, which operates with a random network of recurrent nodes, i.e., nodes with fixed bidirectional coupling, eliminating the need for their training and control. What is trained instead is a weight matrix that is applied to the output of the system. Such a design was implemented with photonic systems~\cite{Kristof2014, Brunner2013}, microwaves~\cite{Torrejon2017} and memristors~\cite{Chao2017}, showing significant recognition rates on a standard set of hand-written digits (Mixed National Institute of Standards and Technology, MNIST). Despite these successfully demonstrations, linear classification algorithms applied directly on the MNIST dataset are still much more precise than any hardware realization of RC.

Following a recent theoretical proposal~\cite{Opala2019}, here we experimentally implement a nonlinear neural network made of a lattice of driven-dissipative condensates of exciton-polaritons to realize a reservoir computer. Exciton-polaritons (hereafter polaritons) are formed in semiconductors (in our case, GaAs quantum wells) placed inside a cavity when excitons hybridize with photons~\cite{RevModPhys.85.299}.
Thanks to the interactions inherited from the electronic component, polaritons possess strong Kerr nonlinearities while keeping their lifetime in the picosecond timescale\cite{Krizhanovskii_UltralowPowerSolitons}. Many experimental milestones have demonstrated successful realisation of proof of principle polariton devices, such as logic gates~\cite{Leyder2007}, switches~\cite{Amo2010}, routers~\cite{Marsault2015} and transistors~\cite{Ballarini2013, Zasedatelev2019} with femtojoule energy cost per operation and few tens of picosecond recovery time~\cite{Dreismann2016}. More recently, lattices of localized polariton condensates have been used to form topological protected states~\cite{Jean2017, Klembt2018} and solving many-body hamiltonians~\cite{Ohadi2017, Berloff2017}. Rather than using polaritons to reproduce binary logic systems, we demonstrate here an alternative neuromorphic information processing architecture and show how the polariton nonlinearity can be harnessed to perform very efficient pattern recognition.

We demonstrate the recognition rate at the level of $93\%$ for the MNIST dataset, higher than in any other hardware implementation of RC to date~\cite{Chao2017,Wang-FPGA-MNIST} and the first to go beyond linear classification for pattern recognition. 
To achieve highly efficient and ultrafast pattern recognition we exploit the exceptional properties of exciton-polaritons with a non-conventional approach to reservoir computing, where data representing a single digit is processed simultaneously instead of sequentially.

We implement the polaritonic reservoir by creating a lattice of $8\times8$ coupled nodes, which act as nonlinear artificial neurons. The transmission intensity of the interacting nodes is the output of the reservoir, which is fed to the final output layer implemented on a computer. Note that this last layer can be implemented all-optically as it requires only a single vector-matrix multiplication at the testing stage, which is a linear operation~\cite{Dolev_DSP}. Although the size of our system is modest, we readily demonstrate a high recognition rate for a MNIST dataset which consists of $28\times28$ handwritten digits. Our result is the first experimental demonstration of MNIST pattern recognition in an optical RC.

\begin{figure}
  \includegraphics[width=\linewidth]{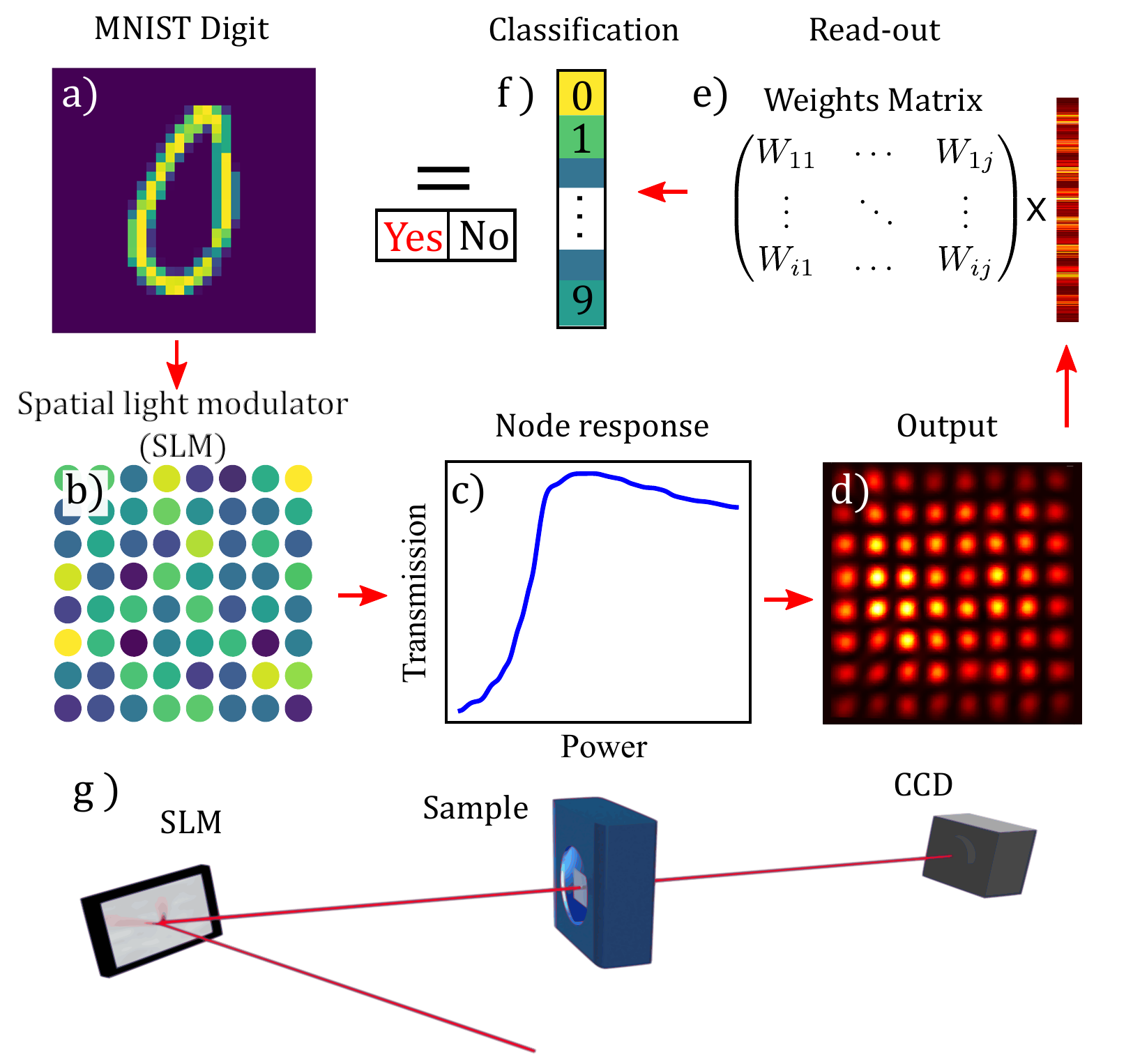}
  \caption{Scheme of the experimental configuration. \textbf{a,} An index $j$ is assigned to each pixel of a 28x28 input image, such that the input intensities are $a_{j}$. These inputs are multiplied by an $8^2\times28^2$ sparse random matrix giving the input as $b_{i} = W_{ij} a_  {j}$. Here $i$ indexes different pixels of an 8x8 image, which directly couples to the 8x8 sites of the reservoir. The same procedure is used for all input images. \textbf{b,} 
The resulting dataset is sent to the SLM to pattern the laser beam. The nonlinear transmission (\textbf{c}) of the polariton RC produces an output (\textbf{d}), which is recorded by a CCD camera and multiplied by the weight matrix (\textbf{e}) to obtain the digit classification (\textbf{f}). \textbf{g,} Scheme of the experimental setup corresponding to the steps described in \textbf{b, c, d}.}
  \label{fig:setup}
\end{figure}

\paragraph*{Results}

The schematic of the experimental procedure is shown in Fig.~1. To distribute the information over the whole reservoir, each digit is converted into a matrix with the same number of elements as the number of nodes in the polariton network. This is done by multiplying the input digit (Fig.~1a) by a random matrix, which is taken the same for all digits. The particular choice of the random matrix has a limited effect on the final recognition efficiency (see Supplemental Information for details on the dataset preparation). Each node in the network is optically created by shaping with a spatial light modulator (SLM) the pump laser beam (wavelength $\lambda=\SI{836}{\nano\meter}$) into an array of spots with controllable intensity and phase. The laser light is injected into a semiconductor microcavity, in which exciton-polaritons are created by tuning the laser frequency at the polariton resonance (see SI). The information of the digit is encoded in the intensities of the spots (Fig.~1b), which excite different realizations of the polariton network, while the junction between neurons is created by imposing a phase difference of $\Delta\theta=\pi$ between adjacent nodes~\cite{Caputo2019}.   
The nonlinearity is provided by the Kerr type interactions between polaritons: for resonant excitation, the curve of the transmission (output) intensity as a function of input power shows a nonlinear behavior (Fig.~1c), which depends on the frequency detuning of the laser beam with respect to the polariton resonance~\cite{Baas2004}. Thus each network node can be considered as an artificial neuron with a nonlinear response function, while the fast polariton propagation within the cavity plane provides the effective connectivity between nodes. The threshold power (of the laser outside the cavity) for each node is typically around few meV, while the fingerprint of the whole polariton RC is limited to only $\SI{150}{\micro\meter}\times\SI{150}{\micro\meter}$. 

The transmission pattern from the polariton lattice (Fig.~1d) is recorded by the camera and the image, sampled with a chosen resolution, is taken as the output of the reservoir. Due to the limited throughput of our SLM, a random choice of 5000 digits ($80\%$ for the training stage, $20\%$ for the testing stage) is used with no preprocessing to evaluate the performance of the network. Training is realized by finding weights in the output layer (Fig.~1e) that minimize the error rate of predictions. 
We implement logistic regression, an efficient algorithm for linear classification, on a computer to adjust the output weights in the readout matrix. In the testing phase, the trained network is used to predict the recognition rate on examples that the system has not seen before. Note that at this stage the computer could be replaced by an all-optical device performing a linear vector-matrix multiplication operation~\cite{Dolev_DSP}. We always compare our results against the pure logistic regression on input data to evaluate the gain of the recognition rate due to the nonlinear transformation performed by the network. 
To this scope, the transfer function of the SLM is calibrated to be strictly linear (see SI), in order to measure solely the contribution of the polariton nonlinearity to the pattern recognition efficiency. This is important because, while the final efficiency can be increased by adding a nonlinear transfer function to the SLM, the polariton array can work at much higher operational speed than the electronically controlled SLM. Indeed, while the SLM is used here as a convenient way to generate an array of polariton nodes, polariton lattices of a few thousands nodes can be realized through lithographic patterning for on-chip integration and this would allow direct excitation with the optical input at repetition rates up to the THz range~\cite{Bloch_EdgeStates}.

\begin{figure}[!t]
  \includegraphics[width=\linewidth]{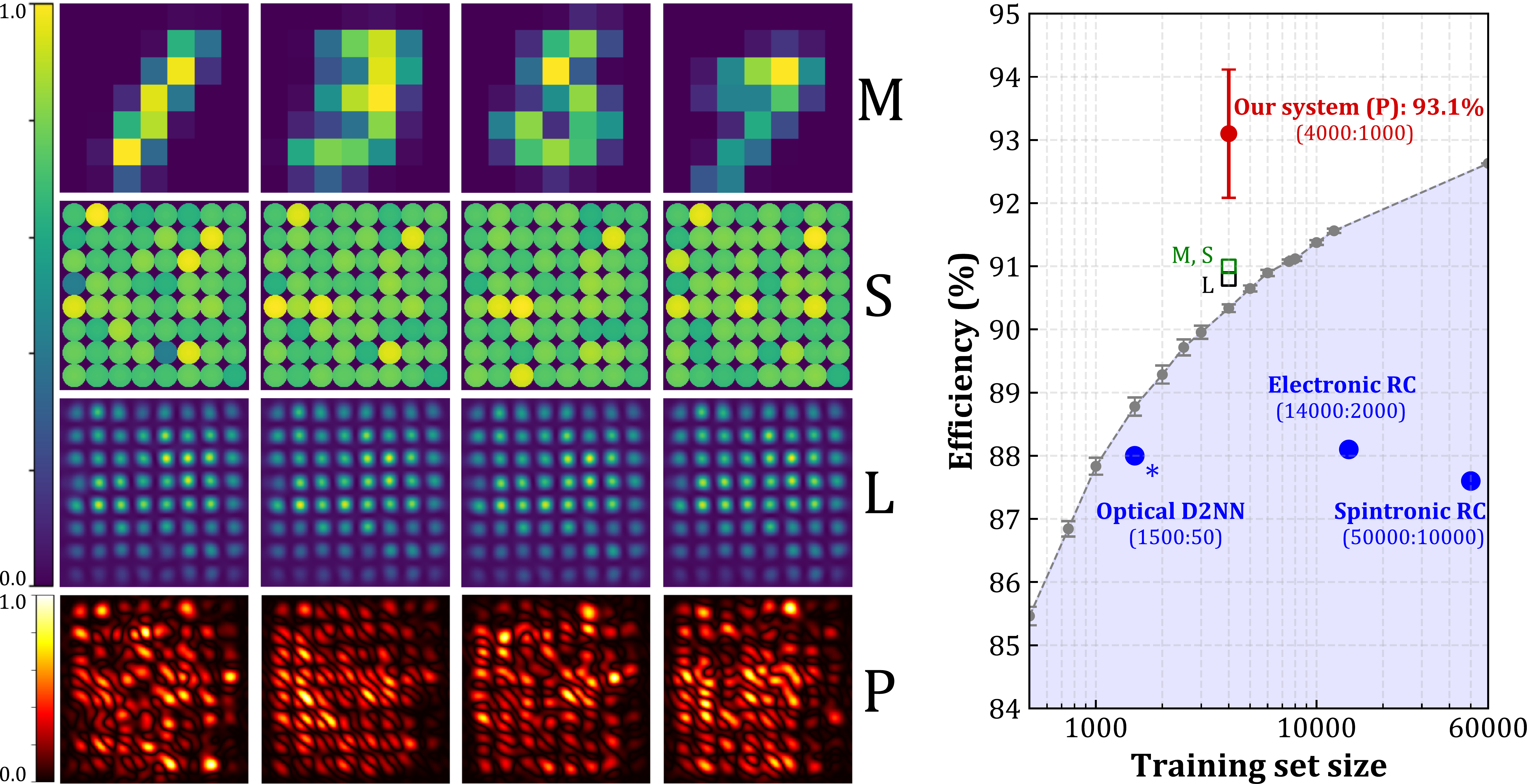}
  \caption{Comparison of polariton RC vs linear classifier. On the left, samples of four digits (1, 3, 5 and 7) processed in the polariton RC. First row ``M'': MNIST digits with 7x7 resolution. Second row ``S'': SLM mask after multiplication of ``M'' by the random matrix. Third row ``L'': laser output after the SLM. Bottom row ``P'': polariton RC, where the quadratic deviation from the mean is shown to highlight the nonlinear redistribution of the information. The spatial scale of the polariton RC is $\SI{150}{\micro\meter}\times\SI{150}{\micro\meter}$. On the right, the success rate obtained with polariton RC is compared to that of linear classifier (shaded region, full MNIST data set of 60000 training digits and 10000 testing digits with 28x28 resolution) and with the state-of-the-art electronic and spintronic RC and terahertz deep neural network (D$^2$NN)~\cite{Chao2017, Lin_D2NN, Jiang2019}. Training and testing set sizes are reported as (training:testing). The ``*'' indicates that in Ref.~\cite{Lin_D2NN} the recognition is measured on a preselection of digits, actually giving an absolute efficiency of $\approx81\%$. The logistic regression recognition rates on the same training and testing sets used in the experiments are shown as black and green open squares for M, S ($91\%$) and L ($90.8\%$), respectively. Logistic regression recognition rate (grey points) are calculated as an average over 10 randomly chosen subsets of the training dataset, and their error bars correspond to the the standard error of the mean. For the experiments, the error bar is calculated by partitioning the testing set of 1000 digits into 10 subsets of 100 digits each as the standard error of the mean recognition rate.}
  \label{fig:err7}
\end{figure}

In Fig.~2, we test the recognition efficiency of the polariton RC on the MNIST dataset. To avoid any spurious effect, such as the possibility that a continuous drift in the temperature would artificially increment the recognition efficiency, the input frames in the dataset are randomly shuffled, confirming the proper operation of the polariton RC in the classification task (see SI). Given the limited reservoir size considered here, we reduce the resolution of the MNIST digits from 28x28 to 7x7 pixels (top row ``M'' in Fig.~2 left). 
Although this operation loses some information, the recognition rate of the polariton RC remains higher than that set by logistic regression, which is illustrated with the shaded grey region in Fig.~2~right for different training set sizes. The exact recognition rate depends on which examples in the MNIST dataset are taken for training and testing. Thus the different data points have error bars, calculated from statistical analysis of different random choices of the training and testing. More training samples results in better performance, but it is remarkable that the polariton RC performs well even with a limited number of training samples. One can also compare to the recognition rate obtained from performing logistic regression on the 7x7 input digit set (point M in Fig. 2 right), performing logistic regression on the 7x7 image SLM mask (point S in Fig. 2 right), and performing logistic regression on the laser beam image after the SLM mask (point L in Fig. 2 right). It can also be seen that the recognition rates obtained in previous hardware implementations of neuromorphic computing were lower.

In Fig.~3, we show that the polariton RC works extremely well also when the resolution of the input image is further reduced to only 4x4 pixels. The logistic regression, performed on the nominal 4x4 input dataset, gives a recognition rate of $81.6\%$. 
Once the signal has been processed by the polariton array, the efficiency increases up to $86.3\%$, showing that polariton nonlinearities allow an increment of $4.7\%$ over the limit set by logistic regression. 
This is possible by repeating the same testing and learning phases with different random matrices, as shown in Fig.~3. The combination of different realizations of the nonlinear array effectively increases the size of the reservoir, allowing to decrease the error rate. In Fig.~3, an improvement of $\approx3\%$ is measured after combining six low-resolution outputs. An equivalent result is obtained on a single repetition by adopting a higher output resolution (see SI). 

\begin{figure}[!t]
  \includegraphics[width=\linewidth]{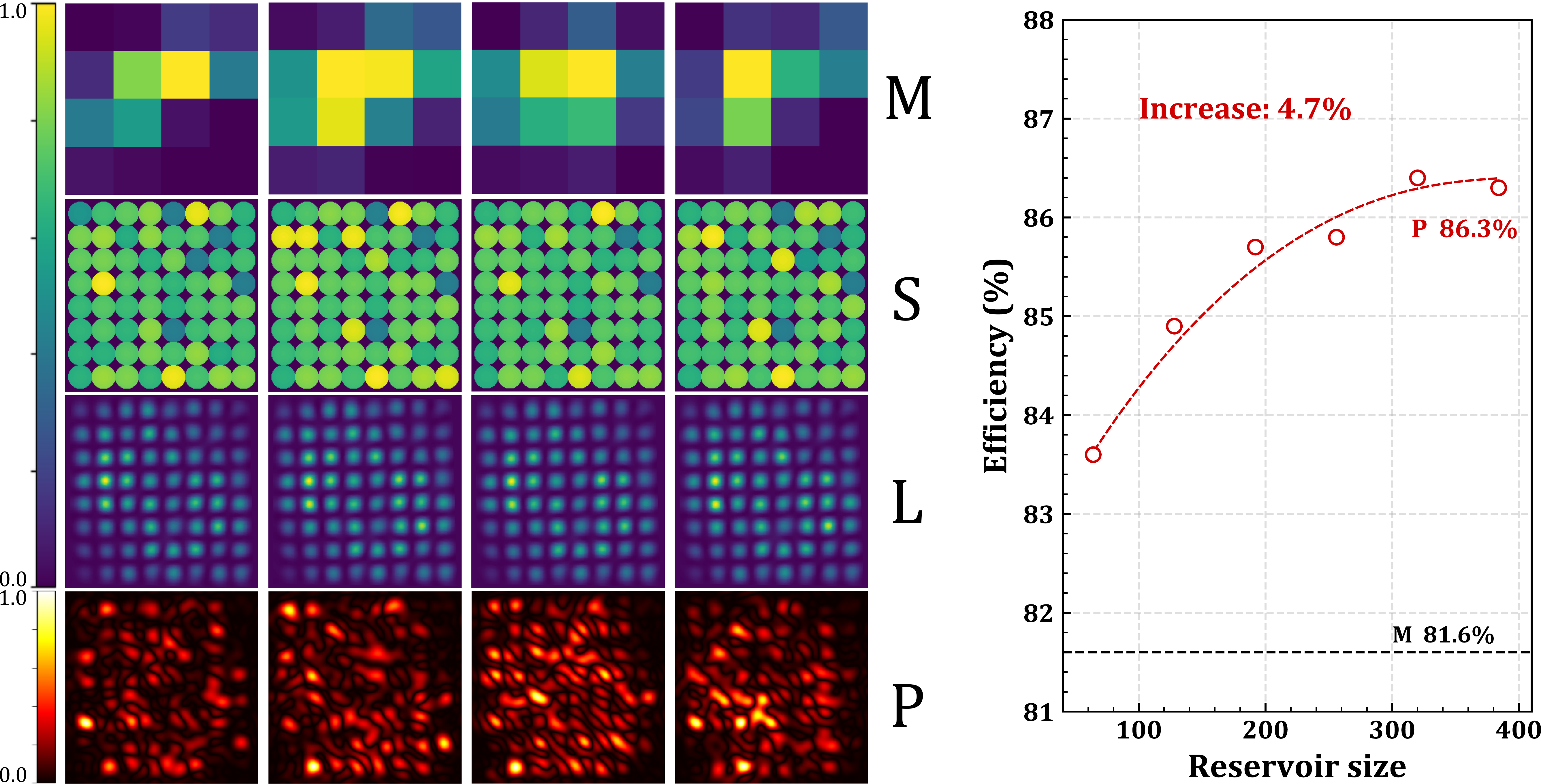}
  \caption{Polariton RC on low resolution images. Linear transformation into a 8x8 matrix is used to distribute the input digit (4x4 resolution) on the whole array by multiplication of the 16 input intensities by a 16x64 matrix. The success rate obtained from logistic regression, both on the 4x4 input digits (samples of 1, 3, 5, 7 input digits are shown in the top row on the left) and on the 8x8 array (second row on the left), is $81.6\%$. The signal after the nonlinear transformation performed by the polariton RC (bottom row on the left, where the quadratic deviation from the mean is shown to highlight the nonlinear redistribution of the information) allows an increase of the recognition rate up to $83.6\%$ already with 64 nodes. The spatial scale of the polariton RC in the bottom row P is $\SI{150}{\micro\meter}\times\SI{150}{\micro\meter}$. On the right, the combination of the information obtained from 6 different random masks allows to increase the effective reservoir size and further enhance the recognition rate to $86.3\%$ (dashed line is a guide to the eye), which is better than logistic regression by a $4.7\%$.}
  \label{fig:err4}
\end{figure}

Finally, we highlight the importance of using hybrid light-matter particles to simultaneously obtain high nonlinearity and efficient connectivity between nodes. Indeed, the nonlinear behaviour of polariton nodes can be simulated by a nonlinear calibration of the SLM transfer function and redirecting its output to the camera. However, even though the success rate is slightly increased over the linear classifier, in this case the result is still lower than that obtained by processing the information in the polariton RC. This is due to the lack of efficient connections between nodes in the SLM array, which are instead present in the polariton network. In Fig.~4, we show that the hopping of polaritons between nodes provides the required connectivity. In Fig.~4b, all the nodes except the central one, marked with A, are pumped using an almost homogeneous SLM pattern. When the intensity of the surrounding nodes is increased, the switched-off node A is activated as well, reflecting the intensity redistribution in the array (blue line in Fig.~4a). Remarkably, connections are not limited to first neighbours but extend farther in the polariton network. In Fig.~4c, also first neighbour nodes A' are switched off and node A can be activated only by the second neighbour nodes (indicated as A'' in Fig.~4). The red line in Fig.~4a shows that nodes A'' contribute to A with a connection strength of about one half as compared to nodes A'.

Moreover, the redistribution dynamics within the network takes place on a timescale comparable to the polariton lifetime. As shown by simulations (see SI), the steady state is reached in this case after about $\SI{50}{\pico\second}$. Even though the present experiments use a continuous wave pump, we anticipate that the recognition rate can be further improved by following the transient dynamics of the system (see the section on time multiplexing in SI). 

\begin{figure}[!t]
  \includegraphics[width=0.9\linewidth,center]{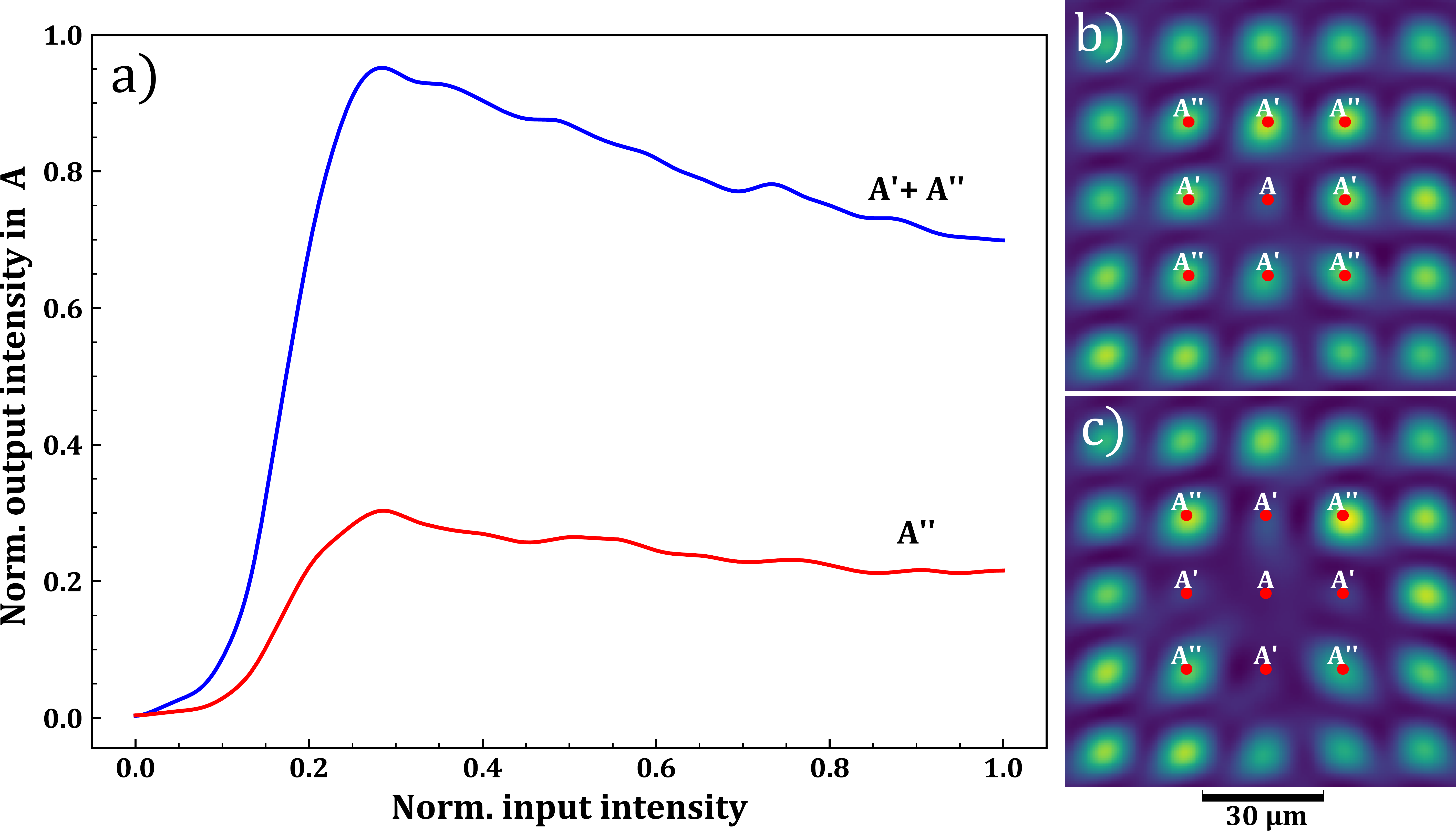}
  \caption{Connectivity between nodes. \textbf{a, }Comparison between the output intensities of node A activated by the contributions from first and second neighbour nodes A' and A'' (blue line), and from second neighbour nodes A'' only (red line). \textbf{b,} Region of the network with pumping power at node A intentionally set to P=0. \textbf{c, } Same as in \textbf{b, } but with power set to P=0 as well at the first neighbour nodes A'. Node A is activated also in this case with a transmitted intensity of about $1/3$ with respect to the configuration shown in \textbf{b}.}
  \label{fig:connex}
\end{figure}

In conclusion, neuromorphic computing in a physical network of multiple connected, nonlinear nodes is demonstrated. 
Up to now, all optical implementations of artificial neural networks have been limited by weak nonlinearities, mostly without overcoming the efficiency of the bare linear classification applied to the raw input data. This lack of fast, energy efficient nonlinearity is one of the most challenging obstacles to be overcome before photon-based neural networks can outperform standard computers. We show that exciton-polaritons are a suited system thanks to their mixed light-matter components: polariton-polariton interactions brings the desired nonlinearities while the photonic component assures the connectivity between nodes and high operational speeds. We measure the recognition efficiency on the MNIST dataset, finding that the error rate is lower than in any other physical implementation used so far. We note that a further increase of the recognition rate can be obtained in polariton networks by exploting the spin degree of freedom and the phase difference between nodes~\cite{Colas2015, Ohadi2016}. With larger polariton arrays, success rates comparable to software implementations of reservoir models can be achieved with present technologies, moving towards a realistic integration of artificial intelligence based on optical systems.



\section*{Acknowledgments}

The ERC "ElecOpteR" grant N. 780757 is acknowledged. S.G. and T. L. were supported by the Ministry of Education (Singapore), grant numbers MOE2017-T2-1-001 and MOE2018-T2-02-068. A.O. acknowledges support from the National Science Center, Poland via Grant No. 2016/22/E/ST3/00045. M.M. acknowledges support from the National Science Center, Poland via Grant No. 2017/25/Z/ST3/03032, under the QuantERA program. We thank R. Houdr\'e for growing the sample.

Data availability: The raw experimental and numerical data used in this study are available from the corresponding author upon reasonable request.

\end{document}